%% file: main.tex
\documentclass[cameraready]{Interspeech}
\title{LRS-VoxMM: A benchmark for in-the-wild audio-visual speech recognition}

\author[]{Doyeop}{Kwak}
\author[]{Jeongsoo}{Choi}
\author[]{Suyeon}{Lee}
\author[]{Joon Son}{Chung}

\address{
    Korea Advanced Institute of Science and Technology, South Korea
}

\email{\{dobbyk, jeongsoo.choi, syl4356, joonson\}@kaist.ac.kr}

\keywords{lip reading, visual speech recognition, audio-visual speech recognition, dataset}

\usepackage{comment}
\usepackage{booktabs}
\usepackage{multirow}
\usepackage{graphicx}
\usepackage{algorithmic}
\usepackage{algorithm}
\usepackage[table]{xcolor}
\usepackage{pifont}
\usepackage{url}
\usepackage{relsize}
\usepackage{hyperref}

\usepackage{fontawesome5}
\usepackage{makecell}
\usepackage{cite}

\begin{document}

\maketitle

\input{section/0_abstract}
\input{section/1_introduction}
\input{section/2_method}
\input{section/3_experiments}
\input{section/4_conclusion}

\bibliographystyle{IEEEtran}
\bibliography{mybib}

\end{document}

%% file: section/0_abstract.tex
\begin{abstract}
We introduce LRS-VoxMM, an in-the-wild benchmark for audio-visual speech recognition (AVSR).  
The benchmark is derived from VoxMM, a dataset of diverse real-world spoken conversations with human-annotated transcriptions. We select AVSR-suitable samples and preprocess them in an LRS-style format for direct use in existing AVSR pipelines.
Compared with commonly used benchmarks, LRS-VoxMM covers a more diverse range of scenarios and acoustic conditions.
We also release distorted evaluation sets with additive noise, reverberation, and bandwidth limitation to support evaluation under severe acoustic degradation. 
Experimental results show that LRS-VoxMM is considerably harder than LRS3 and that the contribution of visual information becomes more evident as the audio signal degrades. LRS-VoxMM supports more realistic AVSR benchmarking and encourages further research on the role of visual information in challenging real-world conditions.
\end{abstract}

%% file: section/1_introduction.tex
\section{Introduction}
Audio-visual speech recognition (AVSR) has advanced substantially in recent years with the development of deep neural models and the availability of large-scale datasets~\cite{shi2022learning, ma2023auto, cappellazzo2025large}. Public benchmarks such as LRW~\cite{chung2016lip}, LRS2~\cite{afouras2018deep}, and especially LRS3~\cite{afouras2018lrs3} have played an important role in this progress by enabling standardized evaluation and reproducible comparison across methods.

At the same time, many widely used benchmarks are drawn from relatively constrained domains such as broadcast news or TED-style talks, where speech is generally clear and recording conditions are comparatively favorable. As audio-only automatic speech recognition (ASR) continues to improve, performance on these benchmarks becomes increasingly saturated. In such settings, the gap between audio-only and audio-visual systems can become relatively small, making it harder to assess when visual information is genuinely necessary and useful. 

More diverse in-the-wild datasets are available~\cite{dahou2023wildvsr,anwar2023muavic,kwak2024voxmm}, but their use as common benchmarks has remained limited in practice. In many cases, this is due to inconsistent data organization, the absence of clearly defined benchmark protocols, or additional preprocessing requirements. As a result, potentially valuable real-world resources remain underutilized for AVSR evaluation. One such resource is VoxMM~\cite{kwak2024voxmm}, an audio-visual dataset of spoken conversations collected from diverse YouTube domains, with rich human-annotated transcriptions and abundant metadata.

In this work, we introduce LRS-VoxMM, a curated benchmark derived from VoxMM and preprocessed in an LRS-style format for direct use in existing AVSR pipelines. By following the conventions of the most widely adopted AVSR benchmark, it enables drop-in use with existing pipelines while providing a more diverse and realistic evaluation setting. We also release synthetically distorted evaluation sets with additive noise, reverberation, and bandwidth limitation for evaluation under severe acoustic degradation. 
Together, these resources support AVSR research in more realistic settings where the contribution of visual information can be more clearly assessed.

%% file: section/2_method.tex
\section{LRS-VoxMM benchmark}
\input{tables/statistics}

VoxMM~\cite{kwak2024voxmm} is a multimodal conversational corpus that provides audio, video, transcripts, speaker labels, face tracks, and other metadata. Unlike conventional audio-visual corpora organized around pre-segmented utterances, it was designed to annotate entire videos as completely as possible, including overlapping speech, off-screen speech, and rich utterance-level metadata. These properties make VoxMM broadly useful for multimodal speech research, but it also means that additional curation is required before it can serve as a standardized AVSR benchmark.

LRS-VoxMM is derived from this corpus by selecting AVSR-suitable samples and preprocessing them to follow the conventions of the LRS family as closely as possible. In particular, we align the input format, transcript style, and directory structure with LRS~\cite{afouras2018deep,afouras2018lrs3} so that the resulting benchmark can be used seamlessly with existing AVSR pipelines built on LRS. All benchmark construction in this work is based on metadata version 1.0.0 of the original release. Further details and download instructions are available on the official project page.\footnote{\url{https://mm.kaist.ac.kr/projects/voxmm}} Table~\ref{tab:dataset_stats} summarizes the statistics of LRS-VoxMM and compares them with existing public AVSR benchmarks.

\noindent \textbf{Preprocessing.}
We use the official VoxMM preprocessing tool\footnote{\url{https://github.com/kaistmm/VoxMM}} with modified configurations for benchmark construction. Audio is resampled to 16\,kHz, and video is processed at 25\,fps with a frame size of $224 \times 224$. Face-track construction and face alignment follow the same procedure as in LRS2/3.

\noindent \textbf{Sample selection.}
We retain utterances suitable for single-speaker AVSR evaluation while keeping the overall sample characteristics close to those of LRS2/3. Specifically, we select samples with utterance durations between 1 and 25 seconds and transcript lengths between 2 and 60 words. Using the metadata, we exclude utterances containing uncertain transcript spans, overlapping speech, or singing. We also remove samples in which the face is only partially visible, the speaker is off-screen, or the visible scene changes within the segment. After automatic filtering, we further conduct manual inspection to remove samples that are unsuitable for benchmark evaluation due to poor data quality. This includes cases where the mouth region is not visible, such as strong side-view faces or head turns, and cases with severe audio-visual synchronization errors.

\noindent \textbf{Transcript normalization.}
We normalize transcripts to follow the LRS2/3 format as closely as possible, since differences in transcript conventions can introduce mismatches in evaluation. Numerical expressions are converted to their actual spoken forms, such as \textit{2026} to \textit{twenty twenty six}. VoxMM annotates a wide range of conversational speech phenomena, including fillers, interjections, and disfluencies. We remove utterances containing disfluencies and retain only filler or interjection forms that are clearly present in LRS2/3, namely \textit{yeah, oh, aha, wow, boom, whoa, hmm, huh, bam, ooh}.

Following the LRS2/3 label format, we additionally provide SyncNet confidence scores and word-level timestamps. Word-level timestamps are obtained by forced alignment using a wav2vec 2.0 LARGE model~\cite{baevski2020wav2vec}. Since VoxMM contains substantially noisier conditions than LRS2/3, we also provide forced-alignment confidence scores to indicate alignment reliability.

\noindent \textbf{Metadata and file structure.}
The directory structure follows LRS2/3, while each sample keeps the original segment index from VoxMM. This allows users to recover utterance-level metadata, such as speaker identity, from the original metadata when needed.

\noindent \textbf{Synthetic distortion sets.}
In addition to the original benchmark split, we release distorted evaluation sets for more severe acoustic conditions. We consider three types of distortion: additive noise, reverberation, and bandwidth limitation. The synthesis procedure largely follows previous work on multi-distortion speech enhancement~\cite{kwak2026ednet}. Additive noise is mixed using the DEMAND noise database~\cite{thiemann2013diverse}. Reverberation is generated using simulated room impulse responses, with room dimensions sampled between 5 and 15\,m for length and width and between 2 and 6\,m for height, and with RT60 sampled between 0.4 and 1.0\,s. Bandwidth limitation is applied using randomly selected low-pass filters, including Butterworth, Bessel, and Chebyshev filters, with cutoff frequencies of 2, 4, or 8\,kHz. Although distorted audio-visual speech recognition has been explored in prior studies~\cite{hong2023watch,rouditchenko2024whisper,kimmulti}, existing evaluations often rely on individually defined distortion scenarios, making direct comparison difficult. By releasing these variants, we aim to provide a standardized benchmark for evaluating AVSR models under distorted acoustic conditions.

We release four distorted splits: \textit{noise\_easy}, \textit{noise\_hard}, \textit{3-dist\_easy}, and \textit{3-dist\_hard}. The \textit{noise} splits contain additive noise only, whereas the \textit{3-dist} splits apply additive noise, reverberation, and bandwidth limitation jointly. The \textit{easy} and \textit{hard} conditions differ only in the SNR range of the additive noise, which is [5, 15]\,dB for the \textit{easy} splits and [-5, 0]\,dB for the \textit{hard} splits. The reverberation and bandwidth-limitation settings are identical across both conditions.

%% file: tables/statistics.tex
\begin{table*}[t]
\centering
\resizebox{1.3\columnwidth}{!}{%

\begin{tabular}{l l l r r r r r}
\toprule
\textbf{Dataset} & \textbf{Source} & \textbf{Split} & \textbf{\# Spk} & \textbf{\# Utt.} & \textbf{Word inst.} & \textbf{Vocab} & \textbf{\# hours} \\
\midrule
LRW      & BBC         & Train-val             & -    & 514k  & 514k  & 500   & 165 \\
         &             & Test                  & -    & 25k   & 25k   & 500   & 8 \\
\midrule
LRS2-BBC & BBC         & Pretrain              & -    & 96k   & 2M    & 41k   & 195 \\
         &             & Train-val             & -    & 47k   & 337k  & 18k   & 29 \\
         &             & Test                  & -    & 1,243 & 6,663 & 1,693 & 0.5 \\
\midrule
LRS3-TED & TED \& TEDx & Pretrain              & 5,090 & 119k  & 3.9M  & 51k   & 407 \\
         &             & Train-val             & 4,006 & 32k   & 358k  & 17k   & 30 \\
         &             & Test                  & 451   & 1,452 & 11k   & 2,136 & 1 \\
\midrule
LRS-VoxMM & 12 Domains & Dev & $\sim$698   & 27k   & 273k  & 14k   & 23.5\\
         &              & Test & $\sim$113   & 2,146   & 22k  & 3,062   & 1.8\\
\bottomrule
\end{tabular}}
\vspace{2mm}
\caption{A comparison of publicly available audio-visual speech recognition datasets. We report the dataset source, split, and the number of speakers, utterances, word instances, vocabulary size, and total duration. LRS-VoxMM is an evaluation-only benchmark derived from the original VoxMM dataset, collected from 12 diverse domains. It consists of one original set and four distorted variants, and its dev/test splits are identical to those of VoxMM.}
\vspace{-6mm}
\label{tab:dataset_stats}
\end{table*}

%% file: section/3_experiments.tex
\section{Experiments}

\subsection{Baselines and checkpoints}

We report results for audio-only ASR, AVSR, and visual speech recognition (VSR) using representative released baselines. Unless otherwise noted, all results are obtained from publicly available official checkpoints under their original training configurations. Since these baselines were trained with different data mixtures, the reported numbers should be interpreted primarily as reference results on LRS-VoxMM rather than as strictly controlled comparisons.

\begin{itemize}[leftmargin=*,topsep=2pt,itemsep=2pt,parsep=0pt]
    \item \textbf{AV-HuBERT~\cite{shi2022learning}.} We use the official AV-HuBERT checkpoints for AVSR and VSR. The released models are pretrained on VoxCeleb2~\cite{chung2018voxceleb2} and fine-tuned on LRS3~\cite{afouras2018lrs3}.
    
    \item \textbf{Auto-AVSR~\cite{ma2023auto}.} We use the official Auto-AVSR checkpoints for AVSR, VSR, and audio-only ASR. These models use LRW-initialized~\cite{chung2016lip} encoders and are trained on LRS2~\cite{afouras2018deep}, LRS3, VoxCeleb2, and AVSpeech~\cite{ephrat2018looking}. Since VoxCeleb2 and AVSpeech do not provide human transcripts for this setting, the training data are augmented with automatically generated pseudo-labels. The original paper considers several public ASR models for pseudo-label generation. We additionally report a reproduced variant trained with the same overall recipe, but without LRS3.

    \item \textbf{Llama-AVSR~\cite{cappellazzo2025large}.} We use the official Llama-AVSR checkpoints for AVSR, VSR, and audio-only ASR. These models are built on pretrained Whisper~\cite{radford2023robust} and AV-HuBERT components, and the released checkpoints are trained on LRS3 or LRS3+VoxCeleb2 depending on the task and configuration.
\end{itemize}

\subsection{Benchmark performance}
\input{tables/main_results}

Table~\ref{tab:avsr_results} shows that all baselines achieve low WER on LRS3 but degrade markedly on the original LRS-VoxMM split. This pattern is consistent across audio-only, audio-visual, and visual-only models, confirming that the benchmark presents a substantially harder evaluation setting than commonly used AVSR benchmarks.

Performance degrades further as distortion increases.  The progression from the original split to \textit{noise\_easy}, \textit{noise\_hard}, \textit{3-dist\_easy}, and \textit{3-dist\_hard} produces a clear monotonic trend for most models, confirming that the synthetic splits provide an effective stress test beyond the already challenging original set.

A key observation is the contrast between audio-only ASR and AVSR. On LRS3, the difference between strong audio-only and audio-visual systems is relatively small, which makes the practical value of the visual modality less clear in an already saturated setting. On the proposed benchmark, however, the gap becomes more apparent. Even on the original split, AVSR models generally outperform their audio-only counterparts. The difference grows further as the acoustic condition becomes more severe: audio-only ASR degrades sharply under \textit{noise\_hard} and especially under the three-distortion setting, whereas AVSR remains substantially more robust. This is consistent with our goal of providing a benchmark in which the contribution of visual information is more clearly exposed under realistic and degraded acoustic conditions.

The reproduced Auto-AVSR$^{*}$ model provides an additional reference point. Removing LRS3 from training leads to the expected drop on LRS3 itself and also causes a modest degradation on the original and \textit{noise\_easy} splits. At the same time, this model is more robust than the official checkpoint in the most severe conditions, achieving lower WER on \textit{3-dist} splits. Although this comparison is not intended as a controlled ablation, it suggests that strong performance on standard benchmarks does not necessarily translate into the best robustness under extreme acoustic degradation.

Experimental results also show that the benchmark is highly challenging for visual-only speech recognition. All evaluated models perform substantially worse than on WildVSR~\cite{dahou2023wildvsr}, which is itself considered a difficult in-the-wild VSR benchmark. Together with the AVSR and audio-only ASR results, this indicates that the difficulty of the benchmark extends to both the acoustic and visual sides.

\subsection{Discussion}

We believe that this difficulty is closely related to the way VoxMM was constructed. Existing AVSR benchmarks are often derived through strong automatic filtering steps designed to retain segments that are easy to align, easy to transcribe, and visually clean. In particular, datasets in the LRS family were filtered using ASR quality criteria during construction, which naturally favors segments on which ASR already performs well. On the visual side, automatic filtering based on lip synchronization confidence also tends to retain well-synchronized and visually clean talking-face segments. As a result, such datasets are highly effective for model development and standardized evaluation, but they may underrepresent the full range of acoustic and visual difficulty encountered in real-world videos.

VoxMM follows a different construction principle. Rather than selecting only segments that pass strict automatic filtering, it was designed to annotate spoken content in full conversational videos as completely as possible. Automatic processing is used where reliable, but segments that are difficult for automatic transcription are still retained through human annotation. This produces a dataset that includes both easy and difficult samples, rather than one consisting primarily of segments that are already favourable to automatic speech recognition. On the acoustic side, this leads to a broader range of recording conditions, speaking styles, background sounds, and utterances that are not necessarily favorable to audio-only recognition. On the visual side, it yields a more realistic distribution of face tracks, including lower resolution, non-frontal views, partial occlusion, mild temporal misalignment, and other imperfect yet natural talking-face segments.

Real-world applications cannot assume that speech is acoustically clean or that visible speaking faces are always frontal, high-resolution, and well synchronized. LRS-VoxMM more directly reflects these conditions and fills an important gap between existing benchmark performance and realistic deployment requirements.

%% file: tables/main_results.tex
\begin{table*}[t]
\centering
\resizebox{0.75\textwidth}{!}{%
\begin{tabular}{cl|c|c|ccccc}
\toprule
& & \textbf{LRS3} & \textbf{WildVSR} & \multicolumn{5}{c}{\textbf{LRS-VoxMM}} \\
\cmidrule(lr){3-3} \cmidrule(lr){4-4} \cmidrule(lr){5-9}
\textbf{Input} & \textbf{Model} & \textbf{original} & \textbf{original} & \textbf{original} & \textbf{noise\_easy} & \textbf{noise\_hard} & \textbf{3-dist\_easy} & \textbf{3-dist\_hard} \\
\midrule
\multirow{2}{*}{A}
& Auto-AVSR      & 0.99 & - & {\scriptsize \textbf{11.07}}\,/\,13.15 & {\scriptsize 13.19}\,/\,13.83 & {\scriptsize 32.59}\,/\,19.12 & {\scriptsize \textbf{43.71}}\,/\,\textbf{49.11} & {\scriptsize \textbf{55.49}}\,/\,\textbf{52.57} \\
& Llama-AVSR     & \textbf{0.75} & - & {\scriptsize 11.78}\,/\,\textbf{11.84} & {\scriptsize \textbf{13.06}}\,/\,\textbf{12.15} & {\scriptsize \textbf{31.99}}\,/\,\textbf{15.92} & {\scriptsize 48.39}\,/\,57.57 & {\scriptsize 68.78}\,/\,64.41 \\
\midrule
\multirow{4}{*}{AV}
& AV-HuBERT         & 1.47 & - & {\scriptsize 18.55}\,/\,20.12 & {\scriptsize 20.06}\,/\,20.96 & {\scriptsize 26.19}\,/\,23.46 & {\scriptsize 33.20}\,/\,38.22 & {\scriptsize 36.25}\,/\,39.22 \\
& Auto-AVSR         & 0.93 & - & {\scriptsize \textbf{8.91}}\,/\,\textbf{11.03} & {\scriptsize \textbf{11.16}}\,/\,11.87 & {\scriptsize 29.38}\,/\,17.19 & {\scriptsize 32.09}\,/\,40.47 & {\scriptsize 38.93}\,/\,42.10 \\
& Auto-AVSR$^{*}$   & 2.21 & - & {\scriptsize 11.75}\,/\,12.90 & {\scriptsize 13.25}\,/\,13.73 & {\scriptsize 23.72}\,/\,17.92 & {\scriptsize 30.06}\,/\,35.16 & {\scriptsize \textbf{34.11}}\,/\,37.10 \\
& Llama-AVSR        & \textbf{0.78} & - & {\scriptsize 11.26}\,/\,11.38 & {\scriptsize 12.35}\,/\,\textbf{11.86} & {\scriptsize \textbf{22.46}}\,/\,\textbf{14.35} & {\scriptsize \textbf{28.93}}\,/\,\textbf{33.95} & {\scriptsize 35.56}\,/\,\textbf{36.62} \\
\midrule
\multirow{4}{*}{V}
& AV-HuBERT         & 27.20 & 51.67 & \multicolumn{5}{c}{{\scriptsize 59.69}\,/\,65.80} \\
& Auto-AVSR         & \textbf{20.61} & 38.36 & \multicolumn{5}{c}{{\scriptsize \textbf{47.36}}\,/\,\textbf{55.15}} \\
& Auto-AVSR$^{*}$   & 27.64 & \textbf{38.26} & \multicolumn{5}{c}{{\scriptsize 49.39}\,/\,56.55} \\
& Llama-AVSR        & 24.31 & 49.22 & \multicolumn{5}{c}{{\scriptsize 62.88}\,/\,70.71} \\
\bottomrule
\end{tabular}}
\vspace{2mm}
\caption{WER (\%) of audio-only (A), audio-visual (AV), and visual-only (V) baselines on LRS3, WildVSR, and LRS-VoxMM with its distorted evaluation splits. All results are obtained from publicly available official checkpoints, except for Auto-AVSR$^{*}$, which denotes our reproduced model trained without LRS3. For LRS-VoxMM, each cell reports WER for the {\scriptsize dev}\,/\,test splits.}
\label{tab:avsr_results}
\vspace{-4mm}
\end{table*}

%% file: section/4_conclusion.tex
\section{Conclusion}
We present LRS-VoxMM, a benchmark-ready in-the-wild AVSR dataset derived from VoxMM and released in a format compatible with existing LRS-based pipelines. By curating AVSR-suitable samples and standardizing the preprocessing, transcript format, and file structure, we make a diverse real-world resource more accessible as a common benchmark. In addition to the original evaluation set, we release distorted variants with additive noise, reverberation, and bandwidth limitation for evaluation under increasingly severe acoustic degradation.

Our results show that this benchmark is substantially more challenging than LRS3 for audio-only, audio-visual, and visual-only speech recognition. They also show that the benefit of visual information becomes more evident as the audio signal degrades, while the visual side remains challenging in its own right due to more realistic and less constrained face tracks. Taken together, these properties make LRS-VoxMM a practical benchmark for studying AVSR beyond already saturated settings and under conditions closer to real-world use.